\documentclass[prb,twocolumn,showpacs,preprintnumbers,amsmath,amssymb,floatfix,superscriptaddress]{revtex4}

\usepackage[]{graphicx}         	
\usepackage{bm}             		
\usepackage{tabularx}

\newcommand{\IEF}{Institut d'Electronique Fondamentale, UMR CNRS 8622, Universit{\'e} Paris-Sud, 91405 Orsay cedex, France}
\newcommand{\IMEC}{IMEC, Kapeldreef 75, B-3001 Leuven, Belgium}

\begin{document}

%
%
\title{Auto-oscillation threshold, narrow spectral lines, and line jitter in spin-torque oscillators based on MgO magnetic tunnel junctions}

\author{T. Devolder}
\affiliation{\IEF}
\author{S. Cornelissen}
\affiliation{\IMEC}
\author{L. Bianchini}
\author{Joo-Von Kim}
\author{P. Crozat}
\author{C. Chappert}
\affiliation{\IEF}
\author{M. {Op de Beeck}}
\affiliation{\IMEC}
\author{L. Lagae}
\affiliation{\IMEC}

\date{\today}

%
%
\begin{abstract}
We demonstrate spin torque induced auto-oscillation in MgO-based magnetic tunnel junctions. At the generation threshold, we observe a strong line narrowing down to 6 MHz at 300K and a dramatic increase in oscillator power, yielding spectrally pure oscillations free of flicker noise. Setting the  synthetic antiferromagnet into autooscillation requires the same current polarity as the one needed to switch the free layer magnetization. The induced auto-oscillations are observed even at zero applied field, which is believed to be the acoustic mode of the synthetic antiferromagnet. While the phase coherence of the auto-oscillation is of the order of microseconds, the power autocorrelation time is of the order of milliseconds and can be strongly influenced by the free layer dynamics.
\end{abstract}


\maketitle

%
%

\section{Introduction}
A spin-polarized current traversing a magnetic multilayer can impart spin angular momentum onto the magnetization, giving rise to an interaction commonly referred to as ``spin-transfer torques" (STT)~\cite{Slonczewski:JMMM:1996,Berger:PRB:1996}. Among the various magnetic systems exhibiting STT effects -- domain walls~\cite{Yamanouchi:Nature:2004}, single films~\cite{Ozyilmaz:PRL:2004}, point contacts~\cite{Rippard:PRL:2004}, vortices~\cite{Pribiag:NatPhys:2007,Mistral:PRL:2008}, granular solids~\cite{Chen:PRL:2006} --, two model systems have been studied more extensively: nanopillars patterned from metallic multilayers~\cite{Albert:APL:2000} or from magnetic tunnel junctions (MTJ)~\cite{Nazarov:APL:2002}. Despite the very different nature of electronic transport in the two systems (i.e. ohmic vs tunneling), it is predicted that spin-torque effect in magnetic tunnel junctions should allow for large-amplitude spin-waves to be excited, well above the thermal occupation of the spin wave populations at equilibrium~\cite{Slavin:ITM:2005}. This phenomenon of current-driven magnetization oscillations is of wide interest for two reasons. First, such spin-torque oscillators reveal some aspects of magnetization dynamics in the presence of a strong frequency nonlinearity, which leads to nontrivial modifications to the excitations. Second, these systems have the potential for applications as compact microwave oscillators for which the frequency is readily tunable with applied magnetic fields and currents.~\cite{Pufall:APL:2005}
	
Using spin-transfer torques to excite magnetization oscillations in spin-valve nanopillars is now a well-established experimental technique.~\cite{Mistral:APL:2006} The signature of genuine auto-oscillation is a clear current threshold behavior~\cite{Kim:PRB:2006,Kim:PRL:2008b} of the microwave signal with three concomitant features: a narrowing of the spectral line, a drastic increase of the line power, and a single mode spectrum with no flicker (1/$f$) noise. Despite intense research activity, auto-oscillations that exhibit all these signatures together are still not commonly observed in tunnel junctions. The debate remains open as to whether the physics of magnetic tunnel junction places some intrinsic restriction on the range in which auto-oscillation can be observed, or rather if the difficulty arises from material limitations, for instance the possibility of fabricating tunnel junctions that can withstand the high current densities required to excite auto-oscillations. In comparison to spin-valves, three physical differences are anticipated in tunnel junctions. First, the spin-transfer torque may include a large component with a field-like symmetry~\cite{Theodonis:PRL:2006}. Second, there may be a nontrivial voltage bias dependence, which would not allow sufficiently large currents to be obtained. Third, the voltage drop across the tunnel barrier may induce significant heating~.\cite{Lee:APL:2008} A number of experiments have been conducted recently with the aim of resolving some of these issues,~\cite{Sankey:NatPhys:2008,Kubota:NatPhys:2008,Nazarov:JAP:2008,Houssamedine:APL:2008} but as far as we can ascertain, none has provided any clear signature of the auto-oscillatory regime.
	
In this article, we present an experimental study of magnetization auto-oscillations in MgO-based tunnel junctions, driven by spin-transfer torques.  We work with tunnel junctions that have been submitted to large current densities in order to stabilize the MgO barrier properties. Upon applying increasing currents, we observe a strong narrowing of the spectral line, down to 6 MHz at an oscillation frequency of 6.06 GHz, which is accompanied by a drastic power increase at the emission threshold, yielding spectrally pure oscillations free of flicker noise. In contrast to the results presented in Ref.~\onlinecite{Houssamedine:APL:2008}, the frequency dispersion of the observed auto-oscilation mode is consistent with that of the acoustic mode of the synthetic antiferromagnet (SAF). Unexpectedly, the electrons transferring spins to the free layer can destabilize both this free layer and the synthetic antiferromagnet for the same current polarity. Finally, we present some evidence of inhomogenous broadening of the spectral line when the linewidth is of the order of a few MHz; we show that the measured linewidth and lineshape is a result of a very narrow line that drifts faster than the accessible capturing times. A spectral line envelope is therefore measured in most cases, giving bounds for the power correlation time and the oscillation phase coherence time.

\section{Samples and experimental methods}
Our nanopillars are fabricated from stacks of composition PtMn (20) / Co$_{70}$Fe$_{30}$  (2) / Ru (0.8) / Co$_{60}$Fe$_{20}$B$_{20}$ (2)  / [Mg 0.9 / natural oxide] / Co$_{60}$Fe$_{20}$B$_{20}$ (3), where figures in parentheses denote film thicknesses given in nm. The films were provided by the SINGULUS company and grown using their Timaris sputter deposition cluster. The devices are rounded rectangles with lateral dimensions of 70$\times$140 nm$^2$ (Fig. 1B, inset). Devices with other dimensions (50$\times$100 and 100$\times$200 nm$^2$) also showed similar auto-oscillations, with minor differences to be reported elsewhere.~\cite{Cornelissen} For a chosen virgin device, the resistances of the P (parallel) and AP (antiparallel) states are 136 and 154 $\Omega$ (magneto-resistance is 13\%). The resistance-area product is 1.4 $\Omega \, \mu \textrm{m}^2$. A current of 1 mA ($10^7$ A/cm$^2$) generates an Oersted field of the order of 1.4 mT at the edge of the nanopillar. In our convention, positive currents describe a flow of electrons from the synthetic antiferromagnet layer to the free layer, which experimentally favors an AP orientation (Fig. 1B). Material fatigue occurs for bias voltages above 0.4 V, after which we observe a large decrease in the junction resistance, with larger currents required to attain threshold for the auto-oscillations. However, we did not observe any qualitative changes to the magnetic properties of the free layer after degradation. 

Following the methods described in Ref.~\onlinecite{Devolder:PRL:2008}, the nanopillars are inserted in series in coplanar waveguide structures using design rules that allow an electrical bandwidth better than 50 GHz.~\cite{Devolder:JAP:2008} The voltage noise spectra were recorded under dc currents with a resolution bandwidth of 2 MHz. For a reason that will become apparent in section IV, we have recorded our spectra by averaging multiple frequency sweeps, except Fig. 3C in which a single sweep curve is presented.

\section{Quasi-static properties}	

The quasi-static properties of a representative virgin nanopillar are presented in Fig. 1.
%
\begin{figure}
\includegraphics[width=8.5cm]{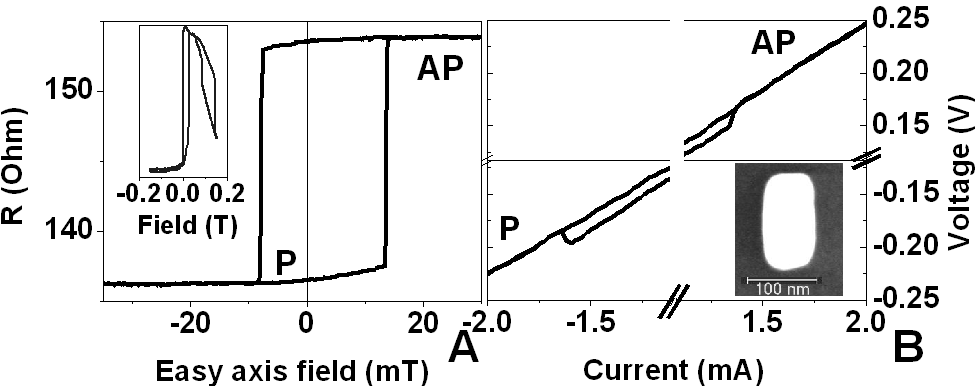}
\caption{Quasi-static properties of the nanopillars in the virgin state. (a) Resistance versus easy axis field minor loop under an applied current of 0.1 mA. Inset: high field curve. (b) Voltage versus applied current curve taken in zero applied field. Inset: scanning electron micrograph of the device after the pillar etching.}
\end{figure}
The resistance versus easy axis field loop (Fig. 1a) indicate uniform magnetization configurations in the remanent states. In our convention, positive fields are defined to favor an antiparallel orientation of the free layer with respect the top layer (reference layer) of the synthetic antiferromagnet (SAF). The SAF layers are not perfectly compensated, which results in a loop shift slightly favoring the P orientation. The Stoner-Wohlfarth astroid (Fig. 2B) confirms the uniaxial anisotropy expected from the device elongated shape. The astroides exhibit the break that is usually present in those size ranges,~\cite{Hayakawa} which is representative of magnetization switching from P to AP through an incoherent micromagnetic path due to the inhomegeneity of the dipole field of the uncompensated SAF. The spin-flop transition of the SAF occurs near 100 to 150 mT depending on the device (Fig. 1a, inset), i.e. in a region sufficiently far from the fields that we will employ to study spin-torque induced auto-oscillations. 

In the virgin state, the voltage versus applied current loops (Fig. 1B )  reveal switching between the two resistance states at critical currents around $\pm$1.5 mA. These switching thresholds increase with sample aging, which is accelerated through the use of large applied currents yielding voltages above 0.4V. The switching currents stabilize typically near 5 mA, provided that the bias never exceeds 8 mA. Expect for Figure 1, we have systematically used this aging procedure to stabilize the junction properties. We emphasize already that these free layer \textit{switching} thresholds are very different from the auto-oscillation threshold. 

The consequence of spin torque can also be observed by looking at the resulting distortion of the Stoner-Wohlfarth astroid of the free layer (Fig. 2).
%
\begin{figure}
\includegraphics[width=8.5cm]{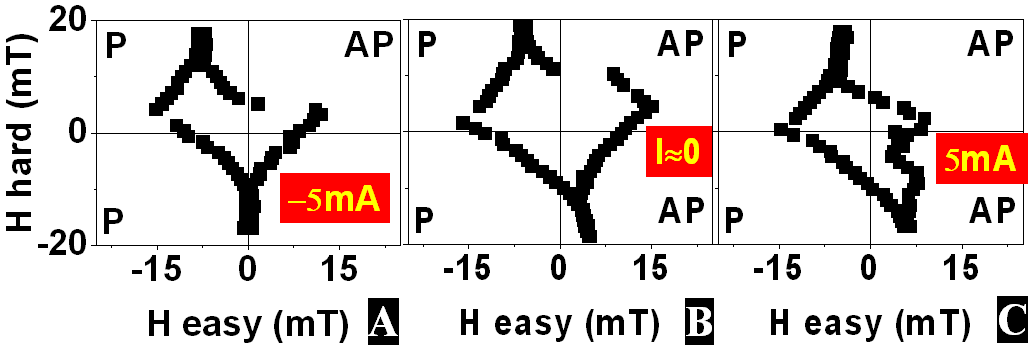}
\caption{(Color online) Stoner-Wohlfarth astroids of a device after the aging procedure using a bias current of 8 mA. A dc current is applied during each astroid measurement. The applied currents are -5 mA, 0.1 mA and 5 mA for the left, central, and right panels, respectively.}
\end{figure}
Under positive currents, the spin-torque favors the AP configuration of the free and reference layers. In addition to a global shrinking which results from current-induced heating, the astroid becomes progressively distorted and a re-entrant border expands near the right corner of the astroid (Fig. 2C). This type of distortion is expected if the spin-torque has a pure Slonczewski form (see Fig. 9 in Ref.~\onlinecite{Sun:PRB:2000}). Under negative currents (Fig. 2a), the current reduces the astroid in all directions by pushing the boundaries toward zero field, recalling that the current significantly heats the junction. In negative current, the astroid distortion is difficult to observe, indicating a lower spin torque per unit current.

\section{Spin torque induced auto-oscillations}
In this section, we identify the current and field regions in which the applied spin-torques effectively favor auto-oscillation of some magnetic degrees of freedom of the nanopillars.  We have analysed the voltage noise under currents up to 8 mA, with easy and hard axis fields up to 0.15 T, including above the spin-flop transition of the SAF. Thermal spin-waves are observed in most regions, but clear auto-oscillation is only observed at low field when the free layer and SAF layers are in the P configuration. In contrast to other studies,~\cite{Nazarov:JAP:2008, Houssamedine:APL:2008} the auto-oscillations we measure are also observed for \emph{zero} applied external fields. Indeed, we believe this to be one of the first experimental demonstrations of field-free current-driven excitations in a magnetic tunnel junction for which a clear auto-oscillation threshold is observed.

Let us now detail how the crossing of the threshold happens and how the auto-oscillations build up (Fig. 3).  
%
%
\begin{figure}
\includegraphics[width=7cm]{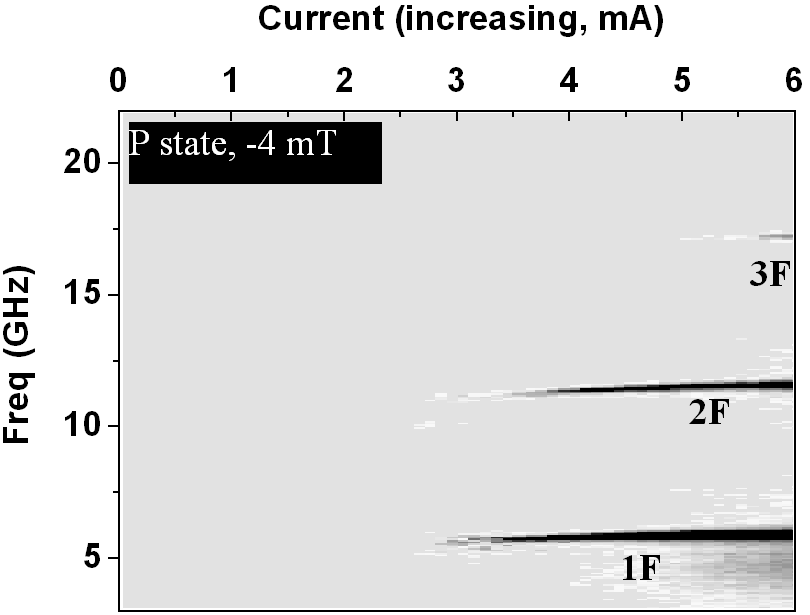}
\caption{Illustration of the current threshold behavior: Voltage noise power spectra (linear scale, arbitrary units) versus current for a junction in the P state at an easy axis field of -4 mT. The device has been submitted to an aging procedure using a bias current of 8 mA.}
\end{figure}
When the current is increased, a narrow line (F or 2F) suddenly emerges from the noise floor slightly below 3 mA. The fundamental frequency (F) of this oscillation at zero field is 6 to 8 GHz, which varies from sample to sample and depends on sample aging. The frequency F is almost independent of the current (Fig. 3) until the oscillation ceases when the ground state of the nanopillar switches to AP slightly below 8 mA (not shown).

In Fig. 4, we present the spectral signature of auto-oscillations in several representative experimental conditions. 
%
\begin{figure}
\includegraphics[width=8.5cm]{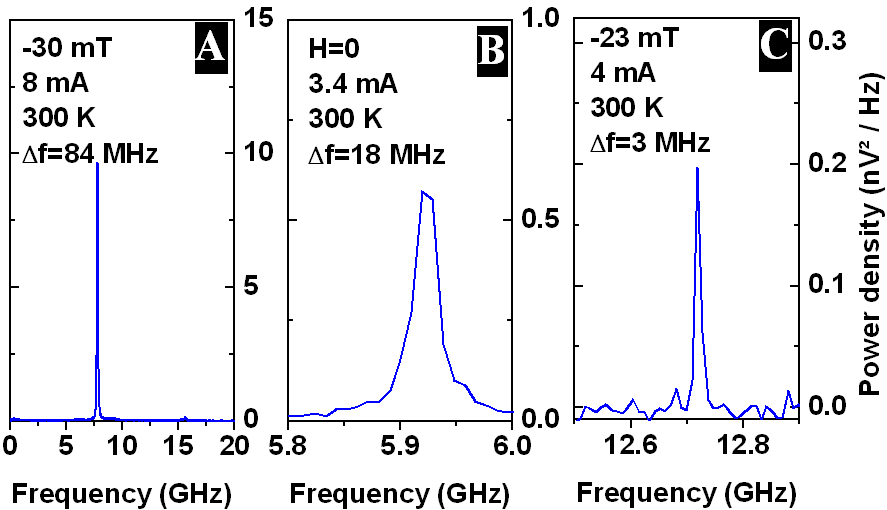}
\caption{(Color online) Voltage noise signature of the autooscillation mode in various experimental conditions for a device after the aging procedure using a bias current of 8 mA.  (a) Spectrum recorded in averaging mode from 0.1 GHz to 20 GHz for easy axis field of -30 mT and 8 mA in the P state. (b) Narrow-band spectrum in averaging mode recorded for zero field and 3.4 mA in the P state. (c) Narrow-band spectrum of the second harmonic (2F) recorded in single sweep mode for an easy axis field of -23 mT and 4 mA in the P state. The quality factor for 2F is $Q=2600$.}
\end{figure}
In Fig. 4a, we present the broadband spectrum across the frequency range of 0.1-20 GHz which shows the absence of any significant $1/f$ noise. Apart from higher-order harmonics, there is no other feature in the whole 0.1 to 20 GHz interval, indicating that the ground state and the excited mode are uniquely defined. In Fig. 4b, a narrower band measurement is presented for an auto-oscillation at exactly zero applied field and under a current right at the emission threshold. Finally in Fig. 4c, we present the narrowest spectral line we have observed at room temperature, which was obtained after optimizing the field magnitude and its orientation in the film plane. The spectral line was obtained using a single-shot measurement  to avoid any envelope effects due to frequency jitter and inhomogeneous line broadening. Note that it is the second harmonic (2F) that is presented in Fig. 4c, as no signal at F was observed above the auto-oscillation threshold at that applied field.

In the auto-oscillation regime, we observed noticeable jitter in the spectral line position. At constant applied currents and fields the measured line shape can vary between each measurement, where the linewidth can fluctuate, for example, from 3 MHz as shown in Fig. 4c, to 18 MHz in a worse case. This fluctuation suggests there is a significant jitter in the oscillation frequency during the acquisition of the spectral data. To get data with statistical significance, the spectra must be recorded in averaging mode, in order to ensure that the line fluctuation translates into a stationary line envelope. Under such measurement constraints, the best measured line envelope width was not better than 6 MHz, corresponding to a quality factor of 1000. Above threshold, the width of the line envelope decreases to a flat minimum in general (Fig. 5), until it increases with line shapes starting to more significantly vary from one spectrum to the next, indicating either unstable ground state or unstable oscillation. 
%
%
\begin{figure}
\includegraphics[width=7cm]{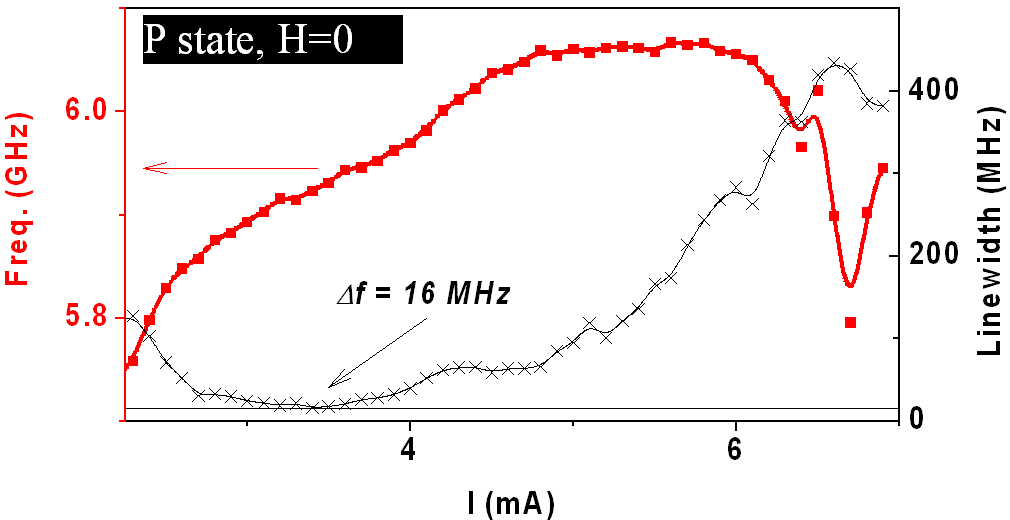}
\caption{(Color online) Frequency and width of spectral line envelope, as a function of applied current at zero applied field. The device has been submitted to an aging procedure using a bias current of 8 mA.}
\end{figure}
%

A prerequisite for observing the auto-oscillations in experiment is an adequate preparation of the initial ground state. In Fig. 6, we present the power spectra in a density map as a function of applied easy axis field for an applied current of 4 and 6 mA, which are above threshold. Both increasing and decreasing branches of the field sweep are presented. The auto-oscillation is observed only in the P state. While the second harmonic (2F) is always present, the fundamental (F) and third harmonic (3F) are only seen in a restricted field interval. We note that the presence of the fundamental F is always observed systematically with the third harmonic 3F (Fig. 6B). We also note that F and 3F are observed only in the field region where the free layer magnetization is not fully saturated (see Fig. 1A). This indicates that the appearance of F and 3F is not related to a change of the oscillation mode, but rather corresponds to a change of the sensitity of the magneto-resistance response to the (unaltered) oscillation mode. The width of the line envelope appears to be independent of the easy-axis field in the monostable P area, i.e. with typical variations between 8 to 19 MHz for fields in the range of -40 to -20 mT (Fig. 6A). At high negative fields, this linewidth floor comes with the absence of the power at F and 3F (Fig. 6B).  

In addition to the auto-oscillation, a faint lower frequency mode with a large linewidth (400 MHz) can generally be detected just before the AP to P (Fig. 6C, mode labelled ``free layer") or the P to AP (not shown) transition. An extrapolation of the frequency of this faint mode indicates that it approaches zero at the anisotropy field of the free layer. This suggests that this mode corresponds to the quasi-uniform spin wave mode of the free layer. The current dependence of the power of this peak (not shown) confirms that this mode is thermally populated rather than being a persistant, spin-torque induced auto-oscillation.

%
%
\begin{figure}
\includegraphics[width=7cm]{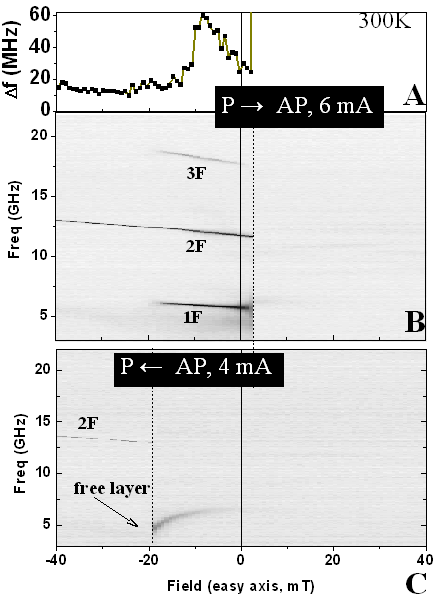}
\caption{Voltage noise properties as a function of the easy axis field and ground state. The device has been submitted to an aging procedure using a bias current of 8 mA. In panels A and B the current is 6 mA and the field is first set negative to prepare the P state. Panel A: line envelope width of the 2F peak. Panel B: noise power densities (log scale). In panel C, the current is 4 mA and the field is first set positive to prepare the AP state.}
\end{figure}
%


To explore further the nature of the excited mode at zero field, we varied the hard axis field at constant applied current and constant easy axis field (Fig. 7). 
%
%
\begin{figure}
\includegraphics[width=7cm]{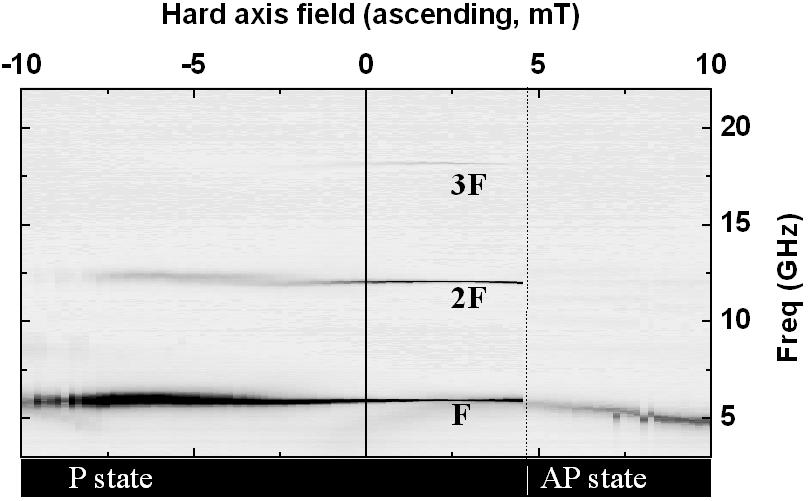}
\caption{Voltage noise spectra (log scale, arbitrary units) as a function of increasing hard axis field for a junction biased with +6 mA and -4 mT along the easy axis at 300K for a junction prepared in the P state. The device has been submitted to an aging procedure using a bias current of 8 mA.}
\end{figure}
When far from the center of the Stoner-Wohlfarth astroid, the width of the line envelope is observed to be very broad (300 MHz), which leads us to speculate whether the observed line corresponds to one mode or several modes with closely-spaced frequencies (see Fig. 8C  for hard axis fields negative and greater than -2 mT). The width of the line envelope reduces to 6 MHz as the hard axis field approaches the middle axis of the astroid (Fig. 8B). Note that this minimum in the  width of the line envelope correlates with a zero slope of the center frequency at the same applied hard axis field. 

This drastic decrease in the width of the line envelope, which in some cases represents a reduction by a factor of 50, is accompanied by a large increase of the peak in the noise power density in such a manner that the total power of the auto-oscillation is almost independent of the hard axis applied field (Fig. 8A). This is another indication that the spectral line we measure corresponds to the envelope of a narrow line, whose power remains constant but suffers an excursion in the frequency within an interval that depends critically on the field. This is in stark contrast to the behavior observed in spin valves in which free layer excitations are measured~\cite{Thadani:PRB:2008}. There, an increase of the hard axis field  leads to a reduction in the linewidth due to an accompanying reduction in the frequency nonlinearity. 
%
%
\begin{figure}
\includegraphics[width=7cm]{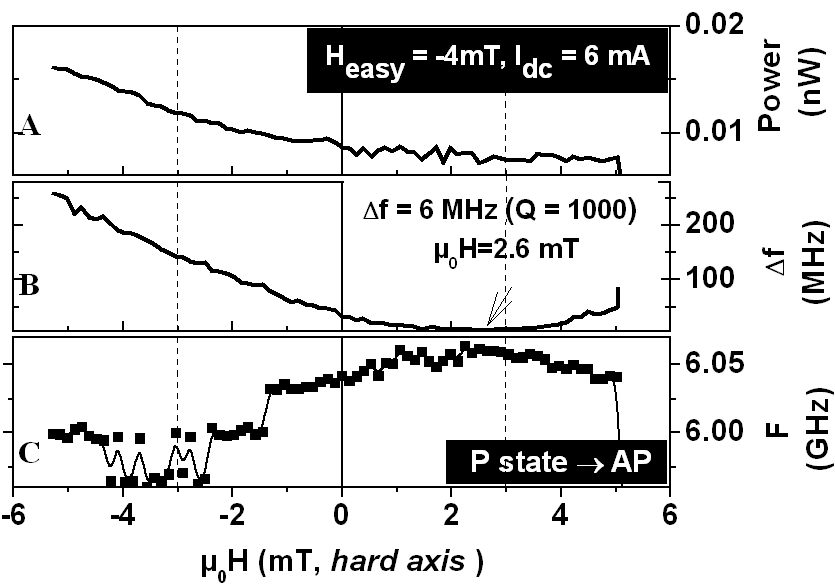}
\caption{Variation of the (a) spectral line power, (b) width of the line envelope, and the (c) frequency, which are extracted from Lorentzian fits to the power spectrum for a junction biased with +6 mA and -4 mT along the easy axis at 300K. The device has been submitted to an aging procedure using a bias current of 8 mA.}
\end{figure}

\section{Discussion}
The auto-oscillation mode we observe at zero applied fields cannot correspond to an eigenmode of the free layer for the following reason. The frequency of a uniform precession mode of the free layer decreases to zero, becoming soft at an applied field along the hard axis that is equal to the shape anisotropy field.  From Fig. 7, it is clear that this is not the case: the mode frequency is almost unchanged when the field approaches the hard axis corners of the astroid at -10 mT. A complementary argument to demonstrate that our auto-oscillation is not a free layer mode can be made using the current dependence of the emission frequency. This dependence exhibits a very small blue shift (Fig. 5), while for a free layer mode there would be a strong red shift for in-plane applied fields, as expected from spin-wave theory.~\cite{Slavin:ITM:2005}

The frequency and its concave dependence on the hard axis field (Fig. 8C) recalls that of the acoustic branch of the SAF eigenexcitations. Reasonable numerical agreement is achieved (not shown) when assuming that the SAF layers have magnetizations of 0.9 T, that they are coupled by interlayer exchange coupling of -2.4x10$^{-4}$ J/m$^2$, and that the exchange pinning strength is 90 mT ~\cite{Liedke:JAP:2006,  Huang:JAP:2006}. Those numbers are based on our experience and are consistent with the experimental spin-flop fields. The interlayer and exchange pinning coupling strengths vary from sample to sample, and decrease with sample aging, probably because of thermal diffusion of Mn or Ru atoms during aging.

Our observations of auto-oscillations under positive currents (electrons flowing from the SAF to the free layer) are unexpected. Indeed, it is well-established for metallic spin-valves that for a dc current that destabilizes the free layer, it stabilizes in turn the reference layer, and vice-versa. This is not the case here: a positive current destabilizes both the free and the SAF layers, by leading to a well-defined auto-oscillation of the synthetic antiferromagnet, but also switches the magnetization of the free layer at a slightly higher current (Figs. 1 and 2). This behavior is suggestive of a non trivial bias dependence of the spin torque in these low resistance-area product magnetic tunnel junctions. One may wonder whether the spin torque solely originates from electrons crossing the oxide (resulting in free layer switching) or if it also partly originates from the diffusive transport in the synthetic antiferromagnet (resulting in SAF auto-oscillations). More elaborate transport calculations, based on realistic stacks in which the magnetization dynamics of the free and synthetic antiferromagnetic layers are taken into account self-consistently, may be required to reproduce these experimental observations. Furthermore, one would also need to account for realistic tunnel barriers, which are more ``permeable'' for electron transport than perfect MgO barriers.


Another important issue concerns the origin of the line fluctuations, keeping in mind that the frequency and oscillator power are intimately related through the strong frequency nonlinearity with the oscillation amplitude.~\cite{Kim:PRB:2006,Tiberkevich:APL:2007,Kim:PRL:2008,Kim:PRL:2008b} From our findings, we shall distinguish two limiting cases: when the free layer is fixed, or when it undergoes substantial fluctuations.  

The first case corresponds to applied fields at the center of the Stoner-Wohlfarth astroid (Fig. 7), or to fields that further stabilize the free layer in a P ground state (Fig. 4C). In those cases, we are usually near a minimum of the width of the line envelope. We have observed frequency fluctuations that were slow enough to be visible from one spectrum measurement to next, while they were fast enough to be seen as random, and not as a gradual drift. This indicates that when the power and frequency deviate from their average, the excess (or default) power relaxes with a characteristic time scale that is comparable to the time needed to record a line in a spectrum, i.e. of the order of some milliseconds with our experimental set-up. In contrast, the phase coherence time of the oscillator is related to the best observed linewidths, and is therefore of the order of a few microseconds. This situation corresponds to the high temperature limit of Ref.~\onlinecite{Tiberkevich:PRB:2008}, in which the measured linewidth appears as an inhomogeneous broadening: power fluctuations are slow enough to result in a modulation of the oscillation frequency. This is corroborated by the fact that when the nonlinearity vanishes (i.e. when experimentaly we have $dF/dI=0$), the frequency is then expected to be immune to power fluctuations and the experimental line envelope width reaches indeed a minimum. 

	In addition, we find that the width of the line envelope increases as the P-to-AP switching field is approached (Fig. 5), or when the free layer magnetization is tilted along the hard axis under a magnetic field (Figs. 7 and 8). This behavior suggests that the free layer dynamics plays an important role on the auto-oscillation of the SAF sub-system. However, present theories ~\cite{Kim:PRB:2006,Kim:PRL:2008b,Tiberkevich:APL:2007} model the linewidth from noise sources that are solely \textit{internal} to the oscillator, and consider this noise to be \textit{white}. In our case, we believe the free layer dynamics is a significant noise source for the SAF oscillations, either as a random field through the dipolar coupling, which would appear as a fluctuation in the oscillation frequency, or as a random modulation of the applied current through a varying spin-transfer efficiency, which would appear as a fluctuation of the supercriticality ($I/I_c$). In particular, the free layer noise is peaked at the frequency of the lowest energy magnon (the main quasi-uniform ferromagnetic resonance mode). Since this mode goes soft when the switching field is approached, its population diverges, thereby contributing a strong colored noise to the torques driving the oscillations in the SAF. This may be the origin of the increase in the observed linewidth near the free layer switching transition.

\section{Conclusion}
In conclusion, we have identified the regions of the field-current parameter space for spin-torque induced auto-oscillations of the synthetic antiferromagnet layers in MgO-based tunnel junctions. The observed oscillation mode exists in zero applied field and is the high amplitude counterpart of the acoustic eigenexcitation of the synthetic antiferromagnet layers. After sample aging, the prerequisites for auto-oscillation include a very low resistance-area product, a ground state in the parallel configuration, and a weak or vanishing hard axis field. Upon increasing the current, three signatures of an auto-oscillation threshold are observed: a drastic spectral line narrowing down to 6 MHz (for an oscillation frequency of 6.06 GHz), a substantial increase in the oscillation power, and a spectrally pure spectrum that is free of $1/f$ noise. In contrast to predictions from simple macrospin models, the auto-oscillation of the synthetic antiferromagnet is observed for a current polarity that also destabilizes the free layer. This results suggests that more complex spin-transport phenomena need to be accounted for in simulations of magnetization dynamics in low resistance-area tunnel junctions made of realistic stacks. Finally, the emission frequency is observed to wiggle randomly during the measurement time, such that the measured line is the envelope of the line frequency excursion. We suggest that noise sources from other magnetic layers in the device, such as the free layer in our case, need to be accounted for in further quantitative comparisons between theory and experiment.

\begin{acknowledgments}
This work was supported by the European Communities programs IST STREP, under Contract No. IST-016939 TUNAMOS, and ``Structuring the ERA'', under Contract No. MRTN-CT-2006-035327 SPINSWITCH, and by the R{\'e}gion Ile-de-France in the framework of C'nano IdF. We thank SINGULUS for providing the films used in this study.
\end{acknowledgments}

\bibliography{articles}

\end{document}